\documentclass[11pt]{article}

\usepackage{graphicx}
\usepackage{epstopdf}
\usepackage{amsmath}
\usepackage{amssymb}
\usepackage{amsfonts}
\usepackage{amsthm}
\usepackage[usenames]{color}
\usepackage{array}
 \usepackage[active]{srcltx}

\usepackage[
      colorlinks=true,
      linkcolor=blue,
      urlcolor=blue,
      filecolor=blue,
      citecolor=red,
      pdfstartview=FitV,
      pdftitle={},
      pdfauthor={},
      pdfsubject={},
      pdfkeywords={},
      pdfpagemode=None,
      bookmarksopen=true
]{hyperref}

\usepackage{epsfig}

\usepackage{hyperref}

\def\beq{\begin{eqnarray}}
\def\eeq{\end{eqnarray}}

\def\be{\begin{equation}}
\def\ee{\end{equation}}
\def\bea{\begin{eqnarray}}
\def\eea{\end{eqnarray}}

\def\be{\begin{equation}}
\def\ee{\end{equation}}
\def\bea{\begin{eqnarray}}
\def\eea{\end{eqnarray}}

\newcommand{\rom}[1]{\mathrm{#1}}

\numberwithin{equation}{section}

\begin{document}

\begin{centering}
 \textbf{\LARGE{The gravitational index of a small black ring}}
     \vspace{0.4in}
    
{\bf  Subhodip Bandyopadhyay$^{1,2}$,
    Gurmeet Singh Punia$^{3,4,5}$     \\ [1mm] Yogesh K. Srivastava$^{1,2}$,    Amitabh Virmani$^{6}$ }

\vspace{0.8cm}

\begin{minipage}{.9\textwidth}\small  \begin{center}
$^1${National Institute of Science Education and Research (NISER), \\ Bhubaneswar, P.O. Jatni, Khurda, Odisha, India 752050}\\
  \vspace{0.5cm}
$^2$Homi Bhabha National Institute, Training School Complex, \\ Anushakti Nagar, Mumbai, India 400085 \\
  \vspace{0.5cm}
$^3$Interdisciplinary Center for Theoretical Study, \\
University of Science and Technology of China, Hefei, Anhui 230026, China \\
\vspace{0.5cm}

$^4$Peng Huanwu Center for Fundamental Theory, Hefei, Anhui 230026, China \\
\vspace{0.5cm}

$^5$School of Science, Huzhou University, Huzhou 313000, Zhejiang, China \\
  \vspace{0.5cm}
$^6$Chennai Mathematical Institute, H1 SIPCOT IT Park, \\ Kelambakkam, Tamil Nadu, India 603103\\
  \vspace{0.5cm}
{\tt \{subhodip.bandyopadhyay, yogeshs\}@niser.ac.in, \\ punia.gurmeet.singh96@gmail.com, avirmani@cmi.ac.in}
\\ $ \, $ \\

\end{center}
\end{minipage}

\begin{abstract}
Certain supersymmetric elementary string states with angular momentum can be viewed as small black rings in a five-dimensional string theory. These black rings have zero area event horizon.  The 4D-5D connection relates these small rings to small black holes  without angular momentum in one less dimension. Recent works have proposed saddle solutions that compute the supersymmetric index for small black holes using gravitational path integral. In this paper, we propose an analogous saddle solution for a  five-dimensional small black ring. The dominant contribution comes from a black ring saddle that rotates in both independent planes in five dimensions and has a finite area event horizon. We also write the saddle solution as a three center Bena-Warner solution. 
\end{abstract}

\end{centering}

\newpage

\tableofcontents

\section{Introduction}

The counting of the microstates of a class of supersymmetric black holes \cite{Strominger:1996sh, Sen:2007qy} is widely regarded as one of the biggest successes of string theory. The  microscopic answer for the logarithm of the supersymmetric index  reproduces the Bekenstein-Hawking entropy of the corresponding supersymmetric black hole in the large charge limit. This matching though quite remarkable, leaves many questions unanswered regarding the relation between the supersymmetric index and the degeneracy. 

 In recent years,  a method has been proposed for computing the index directly on the
gravity side \cite{1810.11442, 2107.09062} using the gravitational path integral. The key idea is to work at a finite temperature and introduce a chemical potential conjugate to the angular momentum that introduces a factor of $(-1)^F$ in the trace computed by the gravitational path integral. This new understanding of the BPS entropy of supersymmetric black holes is non-trivial even in the classical limit. The agreement is not between the entropies computed on the two sides directly, but between the entropy of an extremal black hole and the entropy of a non-extremal black hole plus a term proportional to the angular momentum carried by the black hole \cite{2107.09062}.

In more than four dimensions supersymmetric black holes and the possible non-extremal rotating black hole saddles
exhibit much richer dynamics \cite{Emparan:2001wn, Emparan:2008eg, Elvang:2004rt}. It is most certainly worth while to have more examples of  non-extremal saddle solutions which correctly reproduce the entropy of extremal black holes. In particular, it is very much desirable to have this understanding for supersymmetric black rings \cite{Elvang:2004rt, Emparan:2006mm}.

On the one hand, the discovery of supersymmetric black rings added several new twists to the subject of the microscopic understanding of black holes in string theory~\cite{Emparan:2006mm}. On the other hand, the microscopics of the so-called small black rings \cite{Elvang:2004xi, Iizuka:2005uv, Dabholkar:2005qs, Dabholkar:2006za} is well developed, including a precise definition of the supersymmetric index \cite{Sen:2009bm} for the five dimensional theory obtained by compactifying heterotic string theory on $T^5$.    Given the recent progress on understanding the non-extremal saddle solutions for small black holes \cite{Chowdhury:2024ngg, Chen:2024gmc, Hegde:2024bmb}, it is natural to ask what are the non-extremal saddle solutions for small black rings in heterotic string theory on $T^5$?

A key aim of this paper is to propose non-extremal saddle solutions for a small black ring.  The small black ring that we study is the two-charge black ring of Elvang and Emparan \cite{Elvang:2003mj}. This is the simplest possible supersymmetric black ring. We note that this black ring is slightly different from the small black rings of \cite{Elvang:2004xi, Iizuka:2005uv, Dabholkar:2005qs, Dabholkar:2006za} in that it does not carry an additional parameter for the dipole charge. For purely technical reasons we have not considered the set-up with a dipole charge. In the following,  it will become clear why we have not consider the dipole charge. In the conclusions section, we will comment on  how this parameter can be included.

Our method to build the non-extremal  charged black ring saddles does not differ in essence from the one that gave the two charge Horowitz-Sen black holes \cite{Horowitz:1995tm} used in the construction of the non-extremal saddles for small black holes \cite{Chowdhury:2024ngg, Chen:2024gmc}.\footnote{Note that  Horowitz and Sen \cite{Horowitz:1995tm} do not present their solution as a two charge solution, but as the most general solution in the duality orbit of the two charge solution.}  A natural guess to construct the non-extremal saddles for the Elvang-Emparan small black ring is to add two charges to the \emph{doubly spinning neutral five-dimensional black ring} and finally take an appropriate supersymmetric limit. This is  precisely what we do in the paper. Fortunately, the doubly spinning neutral five-dimensional black ring solution --- the Pomeransky-Sen'kov black ring --- is well understood \cite{Pomeransky:2006bd}.

The analog of the Pomeransky-Sen'kov black ring solution with an additional dipole charge, though known, is much less explored. For this reason we attempted the construction without the dipole charge. In the conclusions section we expand on this point more. 

The rest the paper is organised as follows. In section \ref{sec:black_rings}, we review the non-extremal black rings with single and double rotations. This review is necessary to set the notation.  In section \ref{sec:susy_single}, we take the supersymmetric limit of the singly spinning solution. It  describes the small extremal black ring whose index saddle we propose in this paper. In section \ref{sec:susy_double}, we take the supersymmetric limit of the  doubly spinning two charge solution. It  describes the non-extremal (formally at finite temperature) supersymmetric doubly rotating (complex) saddle solution. In section \ref{sec:properties}, we study some basic properties of the supersymmetric doubly spinning saddle solution.  In section \ref{sec:BW}, we write the saddle solution in the Bena-Warner form and relate our results to the recent discussion in \cite{Boruch:2025qdq}.  We close with a brief discussion in section \ref{sec:conclusions}. In appendix  \ref{app:solution_generating_technique}, for completeness we review the solution generating technique for adding two charges on a five-dimensional metric. 
 In section \ref{app:chiral-null-model}, we write the saddle solution in the chiral null model form. 

\section{Non-extremal two charge black rings}
\label{sec:black_rings}

Elvang and Emparan  in ref.~\cite{Elvang:2003mj} presented a five-dimensional smooth singly spinning non-extremal black ring with two independent charges.  The technique they used for adding the two charges is rather straightforward. It involves the following three steps: 
\begin{enumerate}
\item We start by adding to a vacuum five-dimensional solution of Einstein equations  a flat sixth dimension $z$.  (In the case of ref.~\cite{Elvang:2003mj} the vacuum five-dimensional solution is the neutral Emparan-Reall black ring \cite{Emparan:2001wn}.) A Lorentz boost (with boost parameter $\alpha$) 
gives a solution with linear momentum in the $z$-direction. 
\item We next apply T-duality in the $z$-direction that exchanges the momentum with a fundamental string charge. 
\item  We apply another boost (with boost parameter $\beta$)
in the $z$-direction to get a solution with both charge and momentum. The resulting configuration is a solution to the classical equations of motion of the low-energy NS-NS sector of superstring theory compactified on T$^4$. Compactifying the $z$-direction on a circle we get the five-dimensional solution of interest. 
\end{enumerate}

The technique can be applied to any vacuum five-dimensional solution. In section \ref{subsec:EE}, we review the Elvang-Emparan solution. In section \ref{subsec:PS}, we apply the same technique to the Pomeransky-Sen'kov black ring solution. Later in the paper, in section \ref{subsec:KerrBlackString} we apply the same technique to the five-dimensional Kerr black string.  For completeness, the technique is reviewed  in appendix  \ref{app:solution_generating_technique}.

\subsection{Singly spinning charged black ring}
\label{subsec:EE}
We start with a very brief review of the neutral Emparan-Reall black ring~\cite{Emparan:2001wn} to set the notation. 
\subsubsection{Emparan-Reall black ring} 
The metric takes the form \cite{Emparan:2006mm},
\bea
  \label{neutral}
  ds^2 &=& -\frac{F(y)}{F(x)} \left( dt-C  \,  R\, \frac{1+y}{F(y)} d\psi_{ER}\right)^2 \nonumber \\[1mm]
       && + \frac{R^2}{(x-y)^2} F(x)
       \left[
         -\frac{G(y)}{F(y)} d\psi_{ER}^2
         -\frac{dy^2}{G(y)}
         +\frac{dx^2}{G(x)}
         +\frac{G(x)}{F(x)} d\phi_{ER}^2
       \right] \, ,
\eea
where
\be
  F(\xi) = 1+\lambda \,\xi \, ,
  \qquad
  G(\xi) = (1-\xi^2)(1+\nu \xi) \, ,
  \qquad
  C  = \sqrt{\lambda(\lambda-\nu)\frac{1+\lambda}{1-\lambda}}\,.
\ee
The dimensionless parameters $\lambda$ and $\nu$ take values
$0< \nu \le \lambda <1$. The dimension-full parameter $R$ sets the scale of the solution.
The coordinate ranges are $-1 \le x \le 1$, $-\infty < y  < -1$,
with asymptotic infinity at $x,y=-1$.
Regularity at infinity requires the angular coordinates to have periodicities (to avoid conical singularities at $x = -1$ and $y = -1$ the angular variables must be identified with period),
\be\label{afperiod}
  \Delta\psi_{ER} = \Delta\phi_{ER} = 2\pi \frac{\sqrt{1-\lambda}}{1-\nu}.
\ee
To avoid also a conical singularity at $x = +1$ we must have
\be
\Delta\phi_{ER} = 2\pi \frac{\sqrt{1+\lambda}}{1+\nu}.
\ee
This is compatible with \eqref{afperiod} only if we take the parameter 
$\lambda$ in terms of $\nu$ as
\be
  \label{balance}
  \lambda=\lambda_c \equiv \frac{2\nu}{1+\nu^2}.
\ee
This condition is often called the balancing condition. The name  ``balancing condition'' originates from the intuition that the angular momentum must be tuned so that the centrifugal force balances the tension and self-attraction of the ring. The event horizon is at $y=-1/\nu$. Once the balancing condition is imposed, $\psi = \psi_{ER}  {\sqrt{1+ \nu^2}}$, $\phi = \phi_{ER}  {\sqrt{1+ \nu^2}}$ have canonical periodicity  $2\pi$.

\subsubsection{Elvang-Emparan two-charge black ring}

Following the solution generating procedure outlined  in appendix  \ref{app:solution_generating_technique}, we get the two charge Elvang-Emparan black ring starting with the neutral Emparan-Reall black ring.\footnote{Note that our coordinates and parameters are slightly different from Elvang and Emparan~\cite{Elvang:2003mj}. They use coordinates and parameters different from the original black ring of \cite{Emparan:2001wn}, which in turn are different from the more modern notation \cite{Emparan:2006mm}. Throughout this paper, we use notation of \cite{Emparan:2006mm} for singly spinning black rings. This also means that we cannot simply take equations from Elvang and Emparan's paper; we had to work out the solution from the start.}  The Lagrangian for the theory is reviewed in appendix \ref{app:solution_generating_technique}. The five-dimensional  Einstein frame metric takes the form,
\bea 
    \nonumber
    ds_5^2 &=& 
    - \frac{1}{(h_\alpha h_\beta)^{2/3}}\frac{F(y)}{F(x)} 
    \Bigg( dt  -C  \,  R\, \frac{1+y}{F(y)} c_{\alpha}c_{\beta}\, d\psi_{ER} \Bigg)^2
      \\[2mm]
    && 
    +(h_\alpha  h_\beta)^{1/3} \frac{R^2}{(x-y)^2} F(x)
       \left[
         -\frac{G(y)}{F(y)} d\psi_{ER}^2
         -\frac{dy^2}{G(y)}
         +\frac{dx^2}{G(x)}
         +\frac{G(x)}{F(x)} d\phi_{ER}^2
       \right] \, , \label{EE_metric}
\eea
where we have defined
\be
  h_\alpha(x,y) 
 = 1 + \frac{\lambda(x-y)}{F(x)} s^2_{\alpha}, \qquad
  h_\beta(x,y) 
   = 1 + \frac{\lambda(x-y)}{F(x)} s^2_{\beta},
\ee
and we use the short hand $c_\alpha = \cosh \alpha$, $c_\beta = \cosh \beta$, $s_\alpha = \sinh \alpha$, $s_\beta = \sinh \beta$. The dilaton $\Phi$ and the extra scalar $\chi$ are given as
\be
  e^{-2\Phi} = h_\alpha(x,y) \, ,\qquad 
e^{-\frac{\sqrt{3}}{\sqrt{2}} \chi} = \frac{h_\beta(x,y)}{\sqrt{h_\alpha(x,y)}} \, .
\ee
The gauge fields are 
\begin{align}
A^{(1)}_t &= \frac{(x-y)\lambda }{F(x)h_\beta(x,y) } c_\beta s_\beta,  &
A^{(1)}_{\psi_{ER}} &= \frac{C R(1+y)}{F(x) h_\beta(x,y)} c_\alpha s_\beta, \\
A^{(2)}_t &= \frac{(x-y)\lambda }{F(x)h_\alpha(x,y) } s_\alpha c_\alpha, &
A^{(2)}_{\psi_{ER}} &= \frac{C R(1+y)}{F(x) h_\alpha (x,y)} s_\alpha c_\beta. 
\end{align}
The antisymmetric tensor field has as non-zero components,
\be
B_{\psi_{ER} t} = \frac{CR(1+y) s_\alpha s_\beta }{F(x) h_\alpha(x,y)}.
\ee
The balancing condition remains the same. The event horizon is at $y=-1/\nu$. The charged solution is smooth everywhere on and outside the horizon.

\subsubsection{Physical properties of the Elvang-Emparan black ring} 
 The mass of the black ring and the horizon area are 
\bea
    \label{M}
    M &=& \frac{\pi R^2}{4 G_N} \frac{\lambda}{1-\nu}  (1 + \cosh 2 \alpha + \cosh 2 \beta), \\
   {\cal A_\rom{H}} &=& 8 \pi^2 R^3 \frac{\nu^{3/2}\sqrt{\lambda(1-\lambda^2)}}{(1-\nu)^2(1+\nu)} c_\alpha c_\beta.
    \label{AH}
\eea
The angular momentum and angular velocity in the $\phi$ direction vanish. In the $\psi$ direction, we have
\bea
J_\psi &=& \frac{\pi R^3}{2 G_N}\frac{\sqrt{\lambda (\lambda-\nu)(1+ \lambda)}}{(1-\nu)^2} c_\alpha c_\beta,\\
\Omega_\psi &=& \frac{1}{R c_\alpha c_\beta}\sqrt{\frac{\lambda - \nu}{\lambda(1 + \lambda)}}.
\eea
The  temperature of the horizon is
\be
T = \frac{1}{4\pi R c_\alpha c_\beta} (1+  \nu) \sqrt{\frac{1-\lambda}{\lambda \nu (1+ \lambda)}}.
\ee
Finally, the U(1) charges of the solution can be defined as,
 \be
\textbf{Q}_i= \frac{1}{16 \pi G_N}\int_{S^3~\mathrm{at}~\infty} e^{-2\Phi_i} \star_5 F^{(i)}, \label{def_charges}
 \ee 
 where $\Phi_1 = \frac{\sqrt{2}}{\sqrt{3}}\chi$ and $\Phi_2 = \Phi - \frac{1}{\sqrt{6}} \chi.$ 
They take values,
 \bea \label{Q_1_EE}
\textbf{Q}_1 &=& \frac{\pi R^2}{4 G_N} \frac{\lambda}{1-\nu}\sinh{2\alpha}, \\
\textbf{Q}_2 &=&\frac{\pi R^2}{4 G_N} \frac{\lambda}{1-\nu}\sinh{2\beta}. \label{Q_2_EE}
 \eea
The solution  also carries a dipole charge fixed in terms of the other parameters of the solution. We will discuss the dipole charge when we consider the supersymmetric limit.

\subsection{Doubly spinning charged black ring}
\label{subsec:PS}
Neutral doubly spinning black ring was constructed by Pomeransky and Sen'kov \cite{Pomeransky:2006bd}. 
 We follow slightly different notation from the one introduced in \cite{Pomeransky:2006bd}. This is mainly to facilitate comparison with the black ring solutions discussed in the previous section.  

\subsubsection{Pomeransky-Sen'kov black ring}
Compared to \cite{Pomeransky:2006bd}, we choose mostly plus signature, and exchange 
$\phi \leftrightarrow \psi$ to conform to the notation of the previous section. In \cite{Pomeransky:2006bd}, angles $\psi$ and $\phi$ 
have been rescaled to have canonical periodicity $2\pi$. This can be a potential source of confusion. For this reason we have used $\psi_{ER}$ and $\phi_{ER}$ to denote the  Emparan-Reall coordinates in the previous section. 

The metric functions take a fairly complicated form in the
general case in which the black ring is not in equilibrium, but they
simplify significantly when the balancing condition is  imposed. In all metrics we write for the doubly spinning black rings, the balancing condition is  imposed. Moreover, $\lambda$ in the solution in \cite{Pomeransky:2006bd} has different meaning from the solution above. It roughly corresponds to the parameter $\nu$ above.  $\nu$ in \cite{Pomeransky:2006bd} is the new parameter of the doubly spinning black ring. We replace $(\lambda, \nu)$ of \cite{Pomeransky:2006bd} with $(\nu, \eta)$, respectively. To summarise,
 \bea
 \lambda_{PS} &=& \nu, \\
 \nu_{PS} &=& \eta.
 \eea
 With this notation, when we take $\eta$ to zero we recover the balanced Emparan-Reall black ring. The doubly spinning black ring metric takes the form, 
\bea\label{twospinmetric}
 ds^2&=&-\frac{H(y,x)}{H(x,y)}(dt+\Omega)^2-\frac{F(x,y)}{H(y,x)}d\psi^2
	-2\frac{J(x,y)}{H(y,x)} d\psi d\phi
  	+\frac{F(y,x)}{H(y,x)}d\phi^2\nonumber\\
&& +\frac{2 k^2 H(x,y)}{(x-y)^2(1-\eta)^2}\left(\frac{dx^2}{G(x)}-
\frac{dy^2}{G(y)}\right),
 \eea
with
\bea
 \Omega &=& \omega_\phi d \phi + \omega_\psi d\psi \\ 
 &=& -\frac{2 k \nu \sqrt{(1+\eta )^2-\nu ^2}}{H(y,x)}\bigl[ (1-x^2) y \sqrt{\eta}d\phi
 \nonumber\\
 &&
 +\frac{1+y}{1-\nu +\eta}
 \left(1+\nu -\eta +x^2 y \eta (1-\nu-\eta)+2\eta x(1-y)\right)\,d\psi\bigr],
 \eea
 and the functions $G$, $H$, $J$, $F$ take the form
\bea\label{GHJF_1}
 G(x)&=&(1-x^2)\left(1+\nu x+\eta x^2\right)\,,\\
 H(x,y)&=& 1+\nu^2-\eta^2+2\nu\eta (1-x^2)y
      +2x\nu(1-y^2\eta^2)+ x^2 y^2 \eta(1-\nu^2-\eta^2)\,,\\
 J(x,y)&=&\frac{2 k^2 (1-x^2) (1-y^2) \nu  \sqrt{\eta}}{(x-y) (1-\eta)^2}
 \,\left(1+\nu ^2 -\eta ^2
  + 2 (x+y)  \nu \eta -x y \eta (1-\nu^2 -\eta^2)\right),\\
 F(x,y) &=& \frac{2 k^2}{(x-y)^2 (1-\eta)^2} \Bigl[G(x) (1-y^2)\left[\left((1-\eta)^2-\nu^2\right)
  (1+\eta )+y \nu (1-\nu^2+2 \eta -3 \eta ^2)\right]\nonumber\\
	&&+ G(y) \bigl[2 \nu ^2
    + x \nu ((1-\eta )^2+\nu^2)
   + x^2\left((1-\eta )^2-\nu ^2\right) (1+\eta)
    +x^3\nu(1-
	\nu^2-3\eta^2+2\eta^3)\nonumber\\
   &&- x^4 (1-\eta ) \eta (-1+\nu ^2+\eta^2)\bigr]\Bigr]. \label{GHJF_4}
\eea
In order to recover the metric (\ref{neutral}) together with the condition \eqref{balance}
one must take $\eta\to 0$, identify $R^2= 2k^2 (1+\nu^2)$, and relate $\psi_{ER} = \frac{\psi}{\sqrt{1+ \nu^2}}$, $\phi_{ER} = \frac{\phi}{\sqrt{1+ \nu^2}}$.

In the above metric, note that the parameter $\eta$ appears under the square root sign. Therefore, for the metric to be real $\eta \ge 0$.  The positivity of the mass of the black ring requires $\nu > 0$ and the finiteness requires $1 + \eta - \nu > 0$ (see below). The finiteness of angular momenta requires $\eta < 1$. Furthermore, the parameters $\nu$ and $\eta$ are restricted to 
$\nu\geq
2\sqrt{\eta}$. 
This condition is a Kerr-like bound on the rotation of the $S^2$. To see this, consider the equation for vanishing $G(y)$, which determines the position of the event horizon within the allowed range $-\infty<y<-1$,
\beq\label{twospinhorizon}
 1+\nu y+\eta y^2=0.
\eeq
Imposing that
the roots of (\ref{twospinhorizon}) are real yields the required bound. The regular event horizon is at
\be
y_h = \frac{- \nu + \sqrt{\nu^2 - 4 \eta}}{2 \eta}. \label{location_EH}
\ee
To summarise, the restrictions  on the parameters $\nu, \eta$ are
\bea
\label{twospinrange}
0\leq \eta <1\,,\qquad 2\sqrt{\eta }\leq\nu<1+\eta.
\eea

\subsubsection{Doubly spinning two-charge black ring}
Following the solution generating procedure outlined  in appendix  \ref{app:solution_generating_technique} applied to the Pomeransky-Sen'kov black ring, we get the doubly spinning two-charge black ring. This solution was also considered by Hoskisson \cite{Hoskisson:2008qq}\footnote{Once again the parameters used there are slightly different from what we use. The ADM mass is not computed correctly in \cite{Hoskisson:2008qq}.}. We have, 
\bea\label{twospinmetric_charge}
 ds^2&=&-\frac{1}{(h_\alpha h_\beta)^{2/3}}\frac{H(y,x)}{H(x,y)}(dt+c_\alpha c_\beta \Omega)^2 + (h_\alpha  h_\beta)^{1/3} \frac{2 k^2 H(x,y)}{(x-y)^2(1- \eta)^2}\left(\frac{dx^2}{G(x)}-
\frac{dy^2}{G(y)}\right) \nonumber\\
&&
+ (h_\alpha  h_\beta)^{1/3} \left[-\frac{F(x,y)}{H(y,x)}d\psi^2
	-2   \frac{J(x,y)}{H(y,x)} d\psi d\phi
  	+  \frac{F(y,x)}{H(y,x)}d\phi^2 \right],
 \eea
where we have defined
\be
  h_\alpha(x,y) 
 = c_\alpha^2 - s_\alpha^2 \frac{H(y,x)}{H(x,y)}, \qquad
  h_\beta(x,y) 
   = c_\beta^2 - s_\beta^2 \frac{H(y,x)}{H(x,y)}.
\ee
The dilaton $\Phi$ and the extra scalar $\chi$ are given as
\be \label{dilaton_5D}
  e^{-2\Phi} = h_\alpha(x,y) \, ,\qquad 
e^{-\frac{\sqrt{3}}{\sqrt{2}} \chi} = \frac{h_\beta(x,y)}{\sqrt{h_\alpha(x,y)}} \, .
\ee
The gauge fields are 
\begin{align}
A^{(1)}_t &= \frac{H(x,y) - H(y,x)}{H(x,y) h_\beta(x,y) } c_\beta s_\beta,  &
A^{(2)}_t &= \frac{H(x,y) - H(y,x)}{H(x,y) h_\alpha(x,y) } c_\alpha s_\alpha, \\
A^{(1)}_{\psi } &= -\frac{H(y,x)}{H(x,y) h_\beta(x,y) } \omega_\psi c_\alpha s_\beta, &
A^{(2)}_{\psi } &= -\frac{H(y,x)}{H(x,y) h_\alpha(x,y) } \omega_\psi c_\beta s_\alpha, \\
A^{(1)}_{\phi } &= - \frac{H(y,x)}{H(x,y) h_\beta(x,y) } \omega_\phi c_\alpha s_\beta, &
 A^{(2)}_{\phi } &=- \frac{H(y,x)}{H(x,y) h_\alpha(x,y) } \omega_\phi c_\beta s_\alpha.
\end{align}
The antisymmetric tensor field has as non-zero components,
\bea
B_{\psi t} &=& -\frac{H(y,x)}{H(x,y) h_\alpha(x,y) } \omega_\psi s_\alpha s_\beta, \\
B_{\phi t} &=& -\frac{H(y,x)}{H(x,y) h_\alpha(x,y) } \omega_\phi s_\alpha s_\beta.
\eea
The event horizon is at $y=y_h$ given in \eqref{location_EH}. The charged solution is smooth everywhere on and outside the event horizon. 

\subsubsection{Physical properties of the doubly spinning two-charge black ring
}

The physical parameters, mass $M$, angular momenta $J_{\psi} $, $J_\phi$,
area of the event horizon $\mathcal{A}_H$, angular velocities $\Omega_{\psi} $, $\Omega_\phi$, and the temperature of the horizon $T$  for the solution can be computed following, say, \cite{Pomeransky:2006bd,Astefanesei:2010bm}. We find,
 \bea
 M&=&\frac{k^2 \pi \nu }{G_N (1-\nu +\eta )} \left(1 + \cosh 2 \alpha + \cosh 2 \beta \right), \label{mass} \\
J_\phi&=&\frac{4 k^3 \pi  \nu  \sqrt{\eta } \sqrt{(1+\eta )^2-\nu^2}}{G_N (1-\eta )^2 (1-\nu +\eta)} c_\alpha c_\beta, \label{J-phi} \\
  J_\psi&=&\frac{2 k^3 \pi  \nu  \left(1+\nu -6 \eta +\nu  \eta +\eta ^2\right)
 \sqrt{(1+\eta )^2-\nu ^2}}{G_N (1-\eta )^2 (1-\nu +\eta )^2} c_\alpha c_\beta,  \label{J-psi}  \\
\mathcal{A}_H&=&-\frac{32\pi^2
  k^3(1+\nu+\eta)\nu}{(y_h-1/y_h)(1-\eta)^2} c_\alpha c_\beta, \label{area_two_spin} \\
\Omega_{\phi}&=&\frac{1}{k c_\alpha c_\beta}\frac{\nu (1+ \eta )-(1-\eta )\sqrt{\nu ^2-4 \eta } }{4 \nu \sqrt{\eta} }\sqrt{\frac{1+\eta-\nu}{1+\eta+\nu }} , \label{Omega-phi} \\
\Omega_{\psi}&=&\frac{1}{2 k c_\alpha c_\beta } \sqrt{\frac{1+\eta -\nu }{1+\eta +\nu }}, \\
T &=& \frac{\sqrt{\nu^2-4 \eta } (1-\eta ) (y_h^{-1}-y_h)}{8 \pi k\, \nu  (1+\nu +\eta ) c_\alpha c_\beta}, \label{eq:temp} 
\eea
where $y_h$ is the location of the horizon given in \eqref{location_EH}. The U(1) charges defined via \eqref{def_charges} take values
\bea
\textbf{Q}_1 &=& \frac{k^2 \pi \nu }{G_N (1-\nu +\eta )} \sinh 2 \alpha,\\
\textbf{Q}_2 &=& \frac{k^2 \pi \nu }{G_N (1-\nu +\eta )} \sinh 2 \beta .
\eea
 One can readily check that these expressions go over to the singly spinning case when $\eta \to 0$.  
 \section{Supersymmetric two charge black ring with single rotation}
 \label{sec:susy_single}
 
We use the balancing condition \eqref{balance} to replace $\lambda$ which appears in the singly spinning solution in favour of $\nu$.  The BPS limit is obtained by taking $\nu \to 0, \alpha \to \infty, \beta \to \infty$ such that we keep the charges fixed~\cite{Elvang:2003mj}. In this limit, the factor 
  $
 \frac{\lambda}{1-\nu}$ that features prominently in the mass and charge expressions \eqref{M}, \eqref{Q_1_EE}--\eqref{Q_2_EE} goes to $2 \nu$.  We define,
 \be
2  \nu \sinh 2 \alpha = \frac{Q_1}{R^{2}}, \qquad 2  \nu \sinh 2 \beta = \frac{Q_2}{R^{2}}. \\ 
\ee
These relations imply, 
\be
\alpha = \frac{1}{2} \sinh^{-1} \frac{Q_1}{2R^{2} \nu}, \qquad
\beta = \frac{1}{2} \sinh^{-1} \frac{Q_2}{2R^{2}\nu}, \label{alpha-beta}
\ee
i.e., $\alpha$ and $\beta$ are replaced in favour of $Q_1$ and $Q_2$. After this replacements, we take the $\nu \to 0$ limit.  

In this limit, $F(y) \to 1$, $F(x) \to 1$, and
\be
h_\alpha \to 1 + \frac{Q_1}{2R^2}  (x-y), \qquad
h_\beta \to 1 + \frac{Q_2}{2R^2}  (x-y),
\ee
and 
\be
C c_{\alpha}c_{\beta} \to \frac{1}{2\sqrt{2}R^2} \sqrt{Q_1 Q_2}.
\ee
As a result, metric \eqref{EE_metric} becomes,
\be
 ds_5^2 = - f^{-2}  \Bigg( dt  -   \frac{1}{2\sqrt{2}R} \sqrt{Q_1 Q_2} \, (1+y)  d\psi \Bigg)^2   +f  ds^2_\mathrm{base},
\ee
where $ds^2_\mathrm{base}$ is the four-dimensional flat base space in ring coordinates,
\be
ds^2_\mathrm{base} = \frac{R^2}{(x-y)^2} 
 \left[ (y^2-1) d\psi^2   +\frac{dy^2}{y^2-1}  +\frac{dx^2}{1-x^2}   +(1-x^2) d\phi^2  \right].\label{ds2-base}
\ee
The coordinates $\phi$ and $\psi$ now have canonical periodicity $2\pi$.
The function $f$ takes the form, 
\be \label{func_f}
f^3 = \left(1 + \frac{Q_1}{2R^2}  (x-y)\right)\left(1 + \frac{Q_2}{2R^2}  (x-y)\right).
\ee
The right hand side of eq.~\eqref{func_f} is a product of two harmonic functions, in the same notation as in ref.~\cite[section 5.1]{Emparan:2006mm}, with the identification $Q_3 = 0, q_1 = 0, q_2 = 0, q_3 = \frac{1}{\sqrt{2}R}\sqrt{Q_1 Q_2}.$  Note that, in this solution the dipole charge $q_3$ is fixed in terms of $Q_1$ and $Q_2$; it is not an independent parameter.

To expand a little more on this point about the harmonic functions, let us write four-dimensional flat space in the coordinates 
\be\label{flatr1r2}
ds^2_\mathrm{base}=dr_1^2+r_1^2 d\phi^2+dr_2^2+r_2^2 d\psi^2.
\ee
The transformation 
\be\label{yxr12}
y=-\frac{R^2+r_1^2+r_2^2}{\Sigma}\,,\qquad x=\frac{R^2-r_1^2-r_2^2}{\Sigma}\,, \qquad \Sigma=\sqrt{(r_1^2+r_2^2+R^2)^2-4R^2r_2^2}\,,
\ee
with inverse
\be\label{r12yx}
r_1=R\frac{\sqrt{1-x^2}}{x-y}\,,\qquad r_2=R\frac{\sqrt{y^2-1}}{x-y}\,.
\ee
takes us to the ring coordinates \eqref{ds2-base}. With these transformations at hand, we observe that
\be
\frac{1}{2R^2} (x-y)= \frac{1}{\Sigma}. \label{Sigma}
\ee
The function $\Sigma^{-1}$ solves the Laplace equation on four-dimensional flat space for a ring source at $r_1 = 0$, $r_2 = R$, $0 \le \psi < 2 \pi$.

A key point is that the event horizon area of this BPS solution is zero. To see this, we observe that the prefactor 
\be
\frac{\nu^{3/2}\sqrt{\lambda(1-\lambda^2)}}{(1-\nu)^2(1+\nu)}
\ee  
in \eqref{AH} goes as $\sqrt{2} \nu^2$ in the $\nu \to 0$ limit. Thus, we see that if we keep the mass finite, the area goes to zero. We conclude that upon taking the supersymmetric limit we get a ``small'' black ring with zero horizon area,
\be
S_\mathrm{BPS} = 0.
\ee

 It is well appreciated \cite{Elvang:2003mj} that the supersymmetric solution described above is dual to a D1-D5 supertube \cite{Lunin:2001fv, Lunin:2002bj, Lunin:2002iz}.  In the F1-P duality frame (for heterotic set-up), this system has been called a ``small''  black ring in 5D supergravity \cite{Iizuka:2005uv, Dabholkar:2005qs, Dabholkar:2006za}; it is expected that if we consider higher derivative  corrections to the supergravity action a finite area horizon would emerge.

 \section{The complex saddle solution with two rotations}
 \label{sec:susy_double}
In this section, we propose a complex  solution with two rotations that serves as a saddle for computing the index for the small BPS black ring of the previous section.

To set the context,  we start with a quick review of the key ideas from \cite{1810.11442, 2107.09062} following \cite{Anupam:2023yns}. The rotation group in five dimensions is $SO(4) = SU(2)_L \times SU(2)_R$. A generic black hole in five dimensions rotates in two orthogonal planes with $\psi$ and $\phi$ as the azimuthal angles. For black rings, motivated by the 4D-5D connection \cite{Gaiotto:2005gf, Gaiotto:2005xt}, the third components of $SU(2)_L$ and $SU(2)_R$ angular momenta are identified as \cite{Iizuka:2005uv, Dabholkar:2005qs, Dabholkar:2006za}, 
 \bea
J_{3L} &=& J_\psi,  \\ 
J_{3R} &=& J_\phi.
\eea
The supersymmetric black rings of the previous section have $ J_{3R} = J_\phi = 0$. The relevant index for a supersymmetric black hole that breaks some $SU(2)_L$ invariant supersymmetries is\footnote{A discussion that also includes fermion zero modes can be found in \cite{Anupam:2023yns}. For a more detailed discussion on the definition of an appropriate index for small black rings, see  section 3 of \cite{Sen:2009bm}.}
\be
e^{S_\mathrm{BPS}} = \mbox{Tr}_{\vec  {\textbf Q}, J_{3L}}\left[ (-1)^F\right],
\ee
where the trace is taken over all states carrying fixed charges $\vec {\textbf Q} = \left({\textbf Q}_1, {\textbf Q}_2 \right)$ and $J_{3L}$. The trace is taken over all values of ${\vec J_{3L}}^{\: 2}, J_{3R}, {\vec J_{3R}}^{\: 2}$.

To compute $e^{S_\mathrm{BPS}}$ from the macroscopic side we begin with the gravitational partition function  with boundary conditions appropriate to that of an Euclidean black hole with inverse temperature $\beta$ and angular velocities $\Omega_{L, R}$. Let the entropy of the Euclidean black hole be $S_0$. Going around the Euclidean time $\tau = i t$ gives the periodic identification (see, e.g., \cite{Cassani:2025sim}), 
\be
(\tau, \psi, \phi) \equiv (\tau  + \beta, \psi -i  \beta \, \Omega_\psi, \phi -i  \beta \, \Omega_\phi). 
\ee
The partition function defined  via the gravitational path integral in  this way computes,
\be \label{GC_Z}
Z(\beta, \Omega_\psi, \Omega_\phi, \mu_1, \mu_2) = \mathrm{Tr}\left[ e^{-\beta M + \beta \Omega_{\psi} J_{\psi} + \beta \Omega_\phi J_{\phi} + \beta \mu_1 {\textbf Q}_1 + \beta \mu_2 {\textbf Q}_2} \right],
\ee
where $\mu_1$ and $\mu_2$ are the chemical potentials for the charges ${\textbf Q}_1$ and ${\textbf Q}_2$ respectively and the trace is over all states.\footnote{Our conventions are such that the first law of black hole mechanics takes a more standard form $T dS = dM - \Omega_i dJ_i - \mu_i d{\textbf Q}_i$. }

For computing the index \cite{1810.11442, 2107.09062}, we set the chemical potential $\Omega_\phi$ dual to $J_\phi$ such that, 
\be \label{beta-Omega}
\beta \Omega_\phi =  2 \pi i.
\ee
The partition function \eqref{GC_Z}  can then be related to  the index $e^{S_\mathrm{BPS}}$. In the classical limit, the relation is \cite{Anupam:2023yns},
\be
S_\mathrm{BPS} = S_0 - \beta M + \beta M_\mathrm{BPS} + 2 \pi i J_\phi.
\ee
Here, $M$ denotes the mass of the black hole solution corresponding to the saddle point that contributes to the index. Although, there is no detailed understanding, in all examples known so far, and on physical grounds we expect,
\be
M = M_\mathrm{BPS}. \label{M-saddle}
\ee
In that case, equation  \eqref{S-saddle} becomes,
\be
S_\mathrm{BPS} = S_0 + 2 \pi i J_\phi. \label{S-saddle}
\ee
This relates the entropy of an extremal (zero temperature BPS) black hole $S_\mathrm{BPS}$ to that of a non-extremal (non-zero temperature but supersymmetric) black hole $S_0$. We test both \eqref{M-saddle} and \eqref{S-saddle} for the small black ring of the previous section.

A key intuition for finding the relevant saddle solutions comes from the following observation:  If we identify $y\to -R/r$, $\nu \to 2m/R$ and
$\eta \to a^2/R^2$, which are the correct identifications in the infinite radius limit of the doubly spinning black ring to the boosted Kerr black string,  then equation \eqref{twospinhorizon} becomes the familiar $r^2-2mr+a^2=0$ for the Kerr black hole. Refs.~\cite{Chowdhury:2024ngg, Chen:2024gmc} identify the saddle solutions for small black holes in four dimensions via the analytic continuation $a \to i a$. This suggests that the analytic continuation we need to consider is to negative values of $\eta$. We also replace the parameter $k$ with the parameter $R$ via
\be
k = \frac{R}{\sqrt{2(1+ \nu^2)}}.
\ee
In the $\eta \to 0$ limit, the parameter $R$ matches on the parameter $R$ of the singly spinning solution.

We propose that the saddle solution for the small black ring is obtained by taking the supersymmetric limit\footnote{Although we call this the supersymmetric limit,  one cannot conclude the existence of Killing spinors until the solution is mapped to a Bena-Warner form, which is done later in the paper. With this hindsight we call the $\nu \to 0$ limit the supersymmetric limit. We reserve the phrase ``BPS'' for referring to the singly spinning supersymmetric black ring.} $\nu \to 0$ of the doubly spinning two-charge black ring. We take the supersymmetric limit keeping the temperature and charges fixed. We define,
\be
\alpha = \frac{1}{2} \sinh^{-1} \frac{Q_1}{2R^{2} \nu}, \qquad
\beta = \frac{1}{2} \sinh^{-1} \frac{Q_2}{2R^{2}\nu}, \label{alpha-beta_2}
\ee
and take $\nu \to 0$. Note that, we use the same relations as we used above for the singly spinning black ring \eqref{alpha-beta}. In particular, we have not introduced any $\eta$ dependent factors in \eqref{alpha-beta_2}. This has both advantages and disadvantages. The advantages are that some of the equations below come out simpler. The disadvantage is that 
the canonically defined charges ${\textbf Q}_1$, $ {\textbf Q}_2$ take the form
\begin{align} \label{canonical-charges}
{\textbf Q}_1 &= \frac{\pi }{4 G_N }  \frac{Q_1}{1 + \eta}, & {\textbf Q}_2 &= \frac{\pi }{4 G_N }  \frac{Q_2}{1 + \eta}, 
\end{align}
with an $\eta$ dependent factor. The advantages outnumber this (minor) disadvantage, and we use \eqref{alpha-beta_2} as we take the supersymmetric limit. 

Let us now look at the supersymmetric limit of the ADM mass, cf.~\eqref{mass}. We get, 
\be
M = \frac{\pi }{4 G_N } \left(\frac{Q_1}{1 + \eta} + \frac{Q_2}{1 + \eta}\right).
\ee
Thus,
\be
M = {\textbf Q}_1 + {\textbf Q}_2,
\ee
which confirms~\eqref{M-saddle}, i.e., the saddle solution has the same mass as the BPS black ring in terms of the physical charges ${\textbf Q}_1$,  ${\textbf Q}_2$. Algebraically speaking, the $\eta$ dependent factor needs to be taken into account when expressing ${\textbf Q}_1$,  ${\textbf Q}_2$ in terms of $Q_1$ and $Q_2$. 

  In this limit, the entropy $S_0= \mathcal{A}_H/(4 G_N)$  of the doubly spinning two-charge black ring 
 behaves as, cf.~\eqref{area_two_spin},
\be
S_0 = \frac{  \pi^2  R  
   \sqrt{-\eta }}{\sqrt{2} G_N (1-\eta)^2} \sqrt{Q_1 Q_2}.
\ee
For $-1 < \eta < 0$, this quantity is non-zero and positive.   In the same limit $J_\phi$ becomes, cf.~\eqref{J-phi},
\be
J_\phi = \frac{  \pi  R  
   \sqrt{\eta }}{2\sqrt{2} G_N (1-\eta)^2} \sqrt{Q_1 Q_2}.
\ee
For the branch 
\be
\sqrt{\eta} = i \sqrt{-\eta},
\ee
in the range $-1 < \eta < 0$,  we observe that the  expressions are such that 
\be
S_0 + 2 \pi i J_\phi = 0. 
\ee
This confirms \eqref{S-saddle}. In particular, $J_\phi$ is purely imaginary. We conclude that the proposed saddle solution satisfies both \eqref{M-saddle} and \eqref{S-saddle}. In this limit, \eqref{beta-Omega} is also satisfied.

 The angular momentum $J_\psi$ of the saddle solution is given as, cf.~\eqref{J-psi},
\be
J_\psi = \frac{\pi R (1-6\eta + \eta^2)}{4 \sqrt{2} G_N (1-\eta)^2 (1+ \eta)}\sqrt{Q_1 Q_2} .
\ee
This expression depends on the parameter $\eta$. The inverse temperate $\beta$ takes the form,  
\be
\beta = \frac{\pi \sqrt{Q_1 Q_2}}{\sqrt{2} R(1-\eta)}. \label{temperature}
\ee
Apart from the physical charges ${\textbf Q}_1, {\textbf Q}_2, J_\psi$, the saddle solution has an extra parameter $\eta$ which allows us to adjust the size of the Euclidean time circle. $\Omega_\psi$ in the supersymmetric limit goes to zero as expected for supersymmetric rotating black holes. Unfortunately, $\beta$ does not go to infinity as $\eta \to 0$. A similar issue as was pointed out by Chen, Murthy, and Turiaci in \cite{Chen:2024gmc} (and it was implicit in the analysis of \cite{Chowdhury:2024ngg}) for four-dimensional small black holes. We do not have anything further to add to this point than what was already said in \cite{Chen:2024gmc}, except that unlike \cite{Chowdhury:2024ngg, Chen:2024gmc} expression  \eqref{temperature} has some non-trivial dependence on the $S^2$ rotation parameter $\eta$. More work is required to fully understand small black holes in the four-dimensional context.   The fact that $\beta^{-1} \neq 0$ in general suffices for our considerations in this paper.

\section{The nature of the saddle solution}
\label{sec:properties}
In this section, we present a detailed  analysis of the saddle solution.  Since the area of the small black ring is zero, the saddle solution is expected to have some singularities. We analyse these singular locations first. 

\subsection{Singular locations}
In the supersymmetric limit $\nu \to 0$, the location of the horizon \eqref{location_EH} becomes,
\be
y = y_h = - \frac{1}{\sqrt{-\eta}}. \label{location_EH_BPS}
\ee
For $\eta < 0$ this is in the physically allowed  range of the $y$ coordinate. In the following, it is convenient to take 
\be
\eta = -b^2,
\ee 
then 
\be
y = y_h = -b^{-1},
\ee is the location of the horizon. 
In the $\nu \to 0$ limit, the five-dimensional dilaton \eqref{dilaton_5D} takes the form,
\be
e^{-2 \Phi} = 1+ \frac{Q_1}{2R^2} \frac{(x-y) (1 + b^2  x y)}{
   (1-b^2) \left(1- b^2  x^2
   y^2\right)}.
\ee
At the poles $x= \pm 1$ of the $S^2$ cross-section of the horizon $y_h = -b^{-1}$ the dilaton $e^{-2 \Phi}$ diverges. This in particular means that the solution is singular at the poles of the $S^2$ cross-section of the horizon.  The situation is analogous to the small black holes analysed in ref.~\cite{Chowdhury:2024ngg}. The singular locations are two rings, 
\be \label{singularity}
y = -b^{-1}, \qquad x = \pm 1, \qquad 0\le \psi < 2 \pi. 
\ee
We leave an investigation of the geometry near the singular locations for the future. For the small black holes this geometry was studied in detail in~\cite{Chowdhury:2024ngg}.\footnote{For the small black rings (with one rotation) this analysis of done in great detail in \cite{Dabholkar:2006za}.} Instead, we focus on the nature of the sources that make up the supersymmetric solution.

\subsection{Four-dimensional base space}

In the  $\nu \to 0$ limit, functions \eqref{GHJF_1}--\eqref{GHJF_4} become,
\begin{align}
\label{GHJF_BPS_1}
&G(x)=(1-x^2)\left(1+\eta x^2\right)\,, &
&H(x,y)= (1 -\eta^2) (1+   \eta x^2 y^2)\,, \\
&J(x,y)= 0, &
&F(x,y) = \frac{R^2}{(x-y)^2} (1+\eta) (1-y^2)(1 + \eta x^2) (1 + \eta x^2 y^2).\label{GHJF_BPS_4} 
\end{align}
In particular, 
\be
\frac{H(y,x)}{H(x,y)} =1.
\ee
With this observation, it is clear that the final metric takes the form, 
\be \label{BPS_saddle_main}
 ds_5^2 = - f^{-2}  \left( dt  + \Omega^{(s)} \right)^2   +f  ds^2_\mathrm{base},
\ee
where 
\bea 
ds^2_\mathrm{base} =  \frac{R^2 H(x,y)}{(x-y)^2(1-\eta)^2}\left(\frac{dx^2}{G(x)}-
\frac{dy^2}{G(y)}\right) -\frac{F(x,y)}{H(y,x)}d\psi^2
  	+  \frac{F(y,x)}{H(y,x)}d\phi^2, \label{flat_base}
\eea
and
\bea
 \Omega^{(s)}  &=& \omega_\phi^{(s)} d \phi + \omega_\psi^{(s)} d\psi  \\
&=& 
 - \frac{\left(1-x^2\right) y  
   \sqrt{Q_1 Q_2}}{2 \sqrt{2} (1-\eta) R \left(1 + \eta  x^2 y^2\right)} i \sqrt{-\eta} d\phi 
   \nonumber \\ && - 
\frac{(1+y) \sqrt{Q_1 Q_2}
   \left(1 -\eta  (1 - 2 x + 2 x y - x^2y)-\eta^2 x^2 y\right)}{2
   \sqrt{2} R \left(1-\eta ^2\right) 
   \left(1 + \eta  x^2 y^2\right)} d\psi. 
   \eea
For $- 1 < \eta < 0$, we note that $\Omega^{(s)}$ necessarily has an imaginary part. Since it is a priori not clear what regularity conditions must be imposed on a complex metric, we cannot comment on the smoothness of the metric. It is as good as a complex metric as other metrics that have featured in similar discussions \cite{Cassani:2025sim}.

A simple calculation using Mathematica shows that $ds^2_\mathrm{base}$ is flat space.\footnote{The 4D base space \eqref{flat_base} might be described as  a flat space metric in ellipsoidal ring coordinates: that is, surfaces of constant $y$ correspond to rings with ellipsoidal (rather than spherical) cross-sections at fixed $\psi$. We leave a more detailed investigation of this point too for the future. We thank Roberto Emparan for this comment.} The following coordinate transformation  
\bea \label{r1r2}
r_1^2 &=&\frac{R^2 \left(1-x^2\right) \left(1+ \eta 
   y^2\right)}{(1-\eta ) (x-y)^2}, \\
r_2^2 &=&\frac{R^2 \left(y^2-1\right) \left(1+ \eta 
   x^2\right)}{(1-\eta ) (x-y)^2},
\eea
brings the base space in a standard form,
\be
ds^2 = dr_1^2 + dr_2^2 + r_1^2 d \phi^2 + r_2^2 d\psi^2.
\ee
Using $\eta = - b^2$, we observe that singularities \eqref{singularity} are at  
\be
y = - b^{-1}, \quad x = + 1 \implies r_1 = 0, \quad r_2 = R_+ = \frac{(1-b)}{\sqrt{1+b^2}}R, \label{R-plus}
\ee  
\be
y = - b^{-1},  \quad x = - 1 \implies  r_1 = 0, \quad r_2 = R_- = \frac{(1+ b)}{\sqrt{1+b^2}} R. \label{R-minus}
\ee  
Note that for $b> 0$, $R_+ < R_-$.  

Even though the solution is singular and the metric is complex, the area calculation is unambiguous. We note that at the horizon $y=y_h = -b^{-1}$ the size of the $\phi$ circle vanishes in the four-dimensional base metric, i.e., $r_1 = 0$ from \eqref{r1r2}. Using this observation in \eqref{BPS_saddle_main} and computing the area of the horizon by considering the induced metric on constant $t$ and constant $y$ surface, we find 
\be
(\det g)\bigg{|}_{y = - b^{-1},~t=\mathrm{constant}}^{(x, \phi, \psi)} = - g^{\mathrm{base}}_{xx} \cdot g^{\mathrm{base}}_{\psi \psi } \cdot \left(\omega_\phi^{(s)}\right)^2.
\ee
All factors on the right hand side of this equation are regular.  In particular, singular terms all cancel out. Thus the area of the solution is well defined. It is  expected that the higher derivative corrections in 5D would smoothen out the singular features of the solution. A detailed analysis of these issues is certainly beyond the scope of this paper.

\subsection{Solution in terms of 4D harmonic functions}

 \begin{figure}[t]
\begin{center}
  \includegraphics[width=7cm]{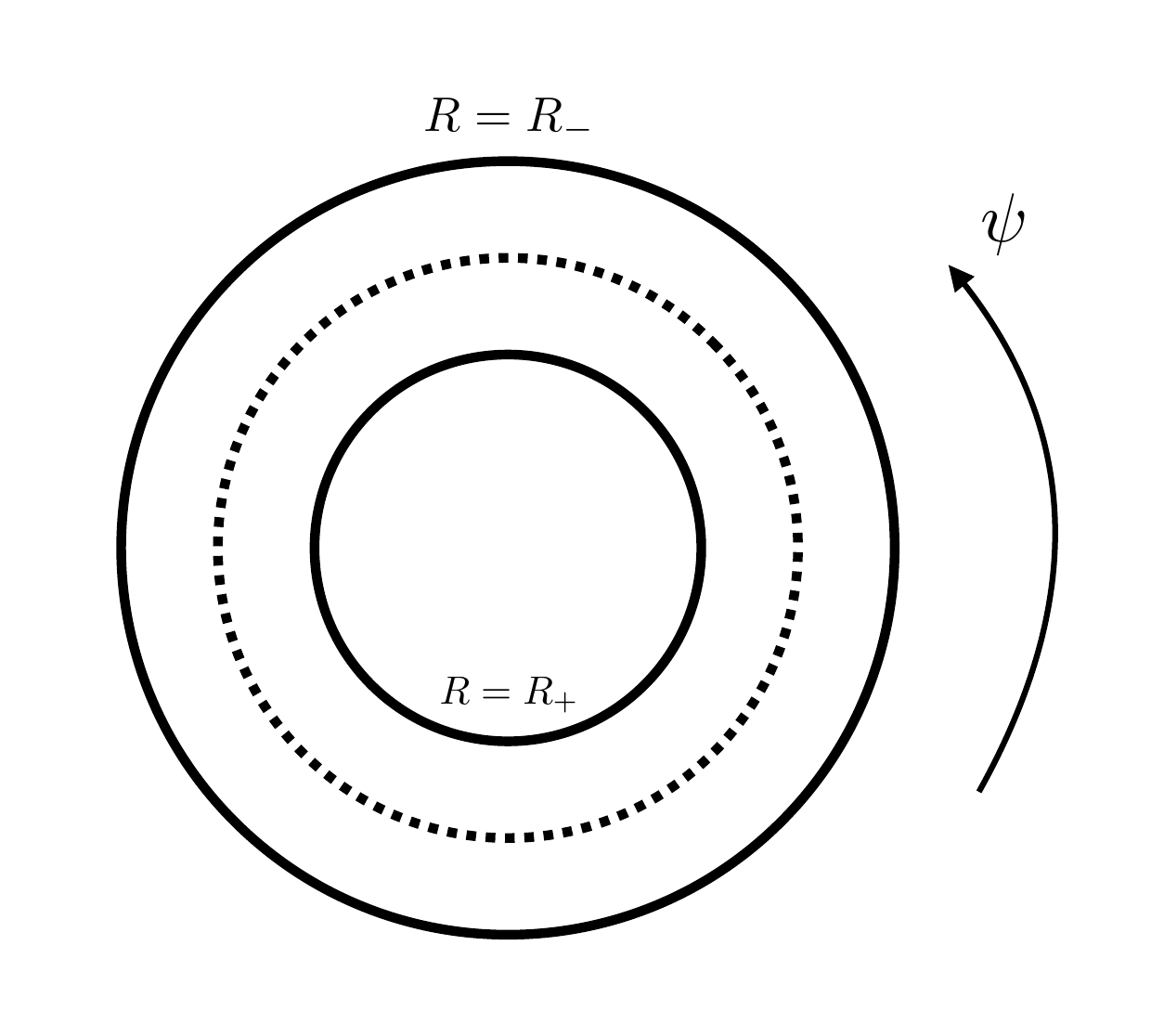}
\end{center}
\caption{\sl For the non-extremal  saddle solution, the  harmonic functions are sourced at two rings $r_1 = 0, r_2 = R_\pm, 0 \le \psi < 2 \pi$ on the four-dimensional flat base space. The $\pm$ signs refer to the poles $x=\pm1$ of the $S^2$ cross-section of the horizon. Since $x=-1$ lies on the outside, $R_- > R_+$. }
\label{fig1}
  \end{figure}

The function $f$ in eq.~\eqref{BPS_saddle_main} is of the form, 
\be
f^3 =h_1 h_2, 
\ee
where the multiplicative factors are obtained via
\bea \label{h1-BPS-x-y}
h_\alpha &\to&  h_1 = 1+ \frac{Q_1}{2R^2} \frac{(x-y) (1 - \eta  x y)}{
   (1+\eta) \left(1+ \eta  x^2
   y^2\right)}, \\
h_\beta &\to&  h_2 =1 + \frac{Q_2}{2R^2} \frac{(x-y) (1 - \eta  x y)}{
   (1+\eta) \left(1+ \eta  x^2
   y^2\right)},\label{h2-BPS-x-y} 
\eea
in the $\nu \to 0$ limit. The vector fields take the following simple form, 
\begin{align} \label{vector-saddle-1}
A^{(1)}_t &= 1 -h_2^{-1},  &
A^{(1)}_{\psi } &=-h_2^{-1} \Omega^{(s)}_{\psi}, &
A^{(1)}_{\phi } &= - h_2^{-1} \Omega^{(s)}_{\phi}, \\
A^{(2)}_t &= 1 -h_1^{-1}, &
A^{(2)}_{\psi } &=- h_1^{-1} \Omega^{(s)}_{\psi}, &
A^{(2)}_{\phi } &=- h_1^{-1} \Omega^{(s)}_{\phi}. \label{vector-saddle-2}
\end{align}
The antisymmetric tensor field has as non-zero components,
\begin{align} \label{B-saddle}
B_{\psi t} &= - h_1^{-1} \Omega^{(s)}_{\psi}, &
B_{\phi t} &= - h_1^{-1} \Omega^{(s)}_{\phi}.
\end{align}

A simple calculation using  Mathematica shows that the function that appears in \eqref{h1-BPS-x-y}--\eqref{h2-BPS-x-y}
\be
\frac{(x-y) (1 - \eta  x y)}{
     \left(1+ \eta  x^2
   y^2\right)} \label{harmonic}
\ee
is a hamonic function for the metric \eqref{flat_base}. The function $f^3$ is thus a product of two harmonic functions. Using  $\eta =- b^2$,
we observe that  the harmonic function \eqref{harmonic} can be split as 
\be
\frac{(x-y) (1 +b^2  x y)}{(1 -b^2 x^2
   y^2)} = \frac{1+b}{2} \left(\frac{x-y}{1- b x y}\right) +  \frac{1- b}{2} \left(\frac{x-y}{1+ b x y}\right),
\ee
such that each of the two terms on the right hand side is harmonic. This split strongly suggests the ``splitting-centers'' picture of  \cite{Boruch:2025qdq}: the harmonic functions of the saddle solution are sourced at two points on a suitable three-dimensional base space. We can make this  precise as follows. 

For a ring source located at $r_1 = 0, r_2 = R$ 
in the $(r_1, \phi, r_2, \psi)$ coordinates,  we saw that the relevant harmonic function is $\Sigma^{-1}$, cf.~\eqref{Sigma}, where
\be
 \Sigma=\sqrt{(r_1^2+r_2^2+R^2)^2-4R^2r_2^2}. 
\ee
For the situation of interest now, the ring sources are located at $r_2 = R_\pm$~\eqref{R-plus}--\eqref{R-minus}. A simple calculation shows that in the $(x, \phi, y, \psi)$ coordinates with metric \eqref{flat_base}, we have,
\bea
\Sigma_+ &=& \sqrt{(r_1^2+r_2^2+R_+^2)^2-4R_+^2r_2^2} = \frac{2R^2 (1-b) (1+ b x y)}{(1+b^2)(x-y)}, \\
\Sigma_- &=& \sqrt{(r_1^2+r_2^2+R_-^2)^2-4R_-^2r_2^2} = \frac{2R^2 (1+b) (1- b x y)}{(1+b^2)(x-y)}.
\eea
Thus indeed,
\bea
h_1 &=& 1 + \frac{Q_1}{2(1+b^2)}\left(\frac{1-b}{1+b}\frac{1}{\Sigma_+} + \frac{1+b}{1-b}\frac{1}{\Sigma_-} \right), \label{h1-BPS}\\
h_2 &=& 1 + \frac{Q_2}{2(1+b^2)}\left(\frac{1-b}{1+b}\frac{1}{\Sigma_+} + \frac{1+b}{1-b}\frac{1}{\Sigma_-} \right), \label{h2-BPS} 
\eea
i.e., the harmonic functions $h_{1,2}$ are sourced at the rings $R_{\pm}$.  These rings are shown in figure~\ref{fig1}. It turns out we can go one step further. The full saddle solution can be written in the Bena-Warner form \cite{Bena:2005va,Bena:2007kg} with \emph{three} centers. This is shown in section \ref{sec:BW}.

\subsection{Infinite radius limit}
\label{subsec:KerrBlackString}
In the infinite radius limit, the black ring saddle solution becomes the saddle solution for a four-dimensional small black hole uplifted to five dimensions with momentum added along the fifth-dimension.  To see this, consider the boosted Kerr black string in five dimensions, 
\bea
d s^2&=& -\left(1-\frac{2 m r \cosh ^2 \sigma}{\Sigma}\right) d t^2+\frac{2 m r \sinh (2 \sigma)}{\Sigma} d t d w+\left(1+\frac{2 m r \sinh ^2 \sigma}{\Sigma}\right) d w^2 \nonumber \\
&& +\frac{\Sigma}{\Delta} d r^2+\Sigma d \theta^2 
 +\frac{\left(r^2+a^2\right)^2-\Delta a^2 \sin ^2 \theta}{\Sigma} \sin ^2 \theta d \phi^2
\nonumber \\
&&-\frac{4 m r \cosh \sigma}{\Sigma} a \sin ^2 \theta d t d \phi- \frac{4 m r \sinh \sigma}{\Sigma} a \sin ^2 \theta d w d \phi ,
\eea
where
\begin{align}
	\Delta&=r^2+a^2-2 m r, &  \Sigma&=r^2+a^2 \cos ^2 \theta,
\end{align}
and where $\sigma$ is the boost parameter,  $m$ the mass parameter, and $a$  is the angular momentum parameters of the 4D Kerr solution. This metric solves vacuum Einstein equations in five-dimensions. Next, consider applying boost-T-duality-boost to add two charges to this black string. In the resulting metric, consider taking the supersymmetric limit as,
\be
m \to 0, \qquad \alpha, \beta \to \infty,
\ee 
such that the charges  
\begin{align}
\widetilde Q_1 &= 4 m \sinh 2 \alpha, & \widetilde Q_2 &= 4 m \sinh 2 \beta,
\end{align}
are kept finite.  Finally, we replace the rotation parameter as $ a = i c$ and set $\sinh \sigma = 1$.
We get, 
\begin{align}\label{black_string_final}
ds^2  &=  - \frac{1}{(\widetilde h_1 \widetilde h_2)^{\frac{2}{3}}} \left(dt + \frac{ (2 \widetilde Q_1\widetilde Q_2)^{\frac{1}{2}} r}{4(r^2 - c^2 \cos^2 \theta)} \left( i c \sin^2 \theta d\phi -  dw \right) \right)^2 + (\widetilde h_1 \widetilde h_2)^{\frac{1}{3}} ds^2_\mathrm{base}, 
\end{align}
where now, 
\be
ds^2_\mathrm{base}  = \frac{r^2 - c^2 \cos^2 \theta}{r^2-c^2} dr^2 + (r^2 - c^2 \cos^2 \theta) d\theta^2 + (r^2 -c^2)\sin^2 \theta d\phi^2 + dw^2 , 
\ee
and
\begin{align}
\widetilde h_1 &= 1 +  \frac{\widetilde Q_1 r }{2(r^2 - c^2 \cos^2\theta)}, &
\widetilde h_2 &= 1 +  \frac{\widetilde Q_2 r}{2(r^2 - c^2 \cos^2\theta)}.
\end{align}
These configurations serve as saddles for computing the gravitational index for a four-dimensional small black hole uplifted to five dimensions with (a specific value of) momentum added along the fifth-dimension.

The black ring saddle metric \eqref{BPS_saddle_main} becomes  \eqref{black_string_final}, upon changing
\begin{align}
x &= \cos \theta, & y &= - \frac{R}{r}, & b &= \frac{c}{R}, &   Q_{1,2} &= R \, \widetilde Q_{1,2}, & \psi &= - \frac{w}{R}. 
\end{align}
in the $R \to \infty$ limit. The other fields also match.

\section{Saddle solution in the Bena-Warner form}
\label{sec:BW}

Using results from \cite{Gauntlett:2004wh}, we can write the harmonic functions 
\eqref{h1-BPS}--\eqref{h2-BPS}  in a form such that they are sourced at two points on a  three-dimensional base space as suggested in \cite{Boruch:2025qdq}. We do the following series of coordinate transformations. First,
\begin{align}
r_1 &= \rho \cos \Theta, &
r_2 &= \rho \sin \Theta,  &
\phi &= \frac{1}{2} \left(\phi_1  + \phi_2\right)&
\psi &= \frac{1}{2} \left(\phi_1  - \phi_2\right),
\end{align}
then,
\begin{align}
\Theta &= \frac{1}{2} \theta, &
\rho &= 2 \sqrt{r},
\end{align}
and finally,
\begin{align}
x_1 &= r \sin \theta \cos \phi_2, &
x_2 &= r \sin \theta \sin \phi_2, &
x_3 &= r \cos \theta.
\end{align}

The four-dimensional flat base space can be written as
\be
ds^2_\mathrm{base} = r (d\phi_1 + \cos \theta d\phi_2)^2 + \frac{1}{r} (dr^2  + r^2 d\theta^2 + r^2 \sin^2 \theta d\phi_2^2), \label{4d-flar-GH}
\ee
In these coordinates, 
\begin{align}
\frac{1}{4}\Sigma_+ &= | \vec x - \vec x_+| = \sqrt{x_1^2 + x^2_2 + \left(x_3 + \frac{R_+^2}{4}\right)^2},  \\
\frac{1}{4}\Sigma_- &= | \vec x - \vec x_-| = \sqrt{x_1^2 + x^2_2 + \left(x_3 + \frac{R_-^2}{4}\right)^2},
\end{align}
where 
\be
\vec x_\pm = \left(0,0, -\frac{R_\pm^2}{4}\right).
\ee 
In the form \eqref{4d-flar-GH}, the three-dimensional base in flat space. Now, we can easily trace through these coordinate transformations and write the full solution in the Bena-Warner form \cite{Bena:2005va,Bena:2007kg}. For the Anupam-Chowdhury-Sen saddle solutions \cite{Anupam:2023yns} this was done in \cite{Hegde:2023jmp, Adhikari:2024zif}. We refer the reader to \cite{Adhikari:2024zif} for a concise review of the Bena-Warner form and its relation to a standard $N=2, D=4$ supergravity notation. 

For the black ring saddles, the eight Bena-Warner functions take the following form. The function  $V$ is simply,
\be
V = \frac{1}{|\vec x|},
\ee
which corresponds to the center at $\vec x = 0$. This is expected, as the four-dimensional base space is flat space. The remaining functions are all either constants (zero or one) or have sources at $\vec x = \vec x_\pm$. Thus clearly, the $\vec x = 0$ center is smooth. The three magnetic functions are 
\bea
K^1 &=& 0, \\
K^2 &=& 0, \\
K^3 &=& \frac{k^3_+}{|\vec x - \vec x_+|} + \frac{k^3_-}{|\vec x - \vec x_-|},
\eea
with
\bea
k^3_+ &=& \frac{\sqrt{Q_1 Q_2}}{4 \sqrt{2}(1+b^2)R} \left( \frac{1-b}{1+b} + i\right),\\
k^3_- &=&  \frac{\sqrt{Q_1 Q_2}}{4 \sqrt{2}(1+b^2)R} \left( \frac{1+b}{1-b} - i\right).
\eea
Note that these coefficients have imaginary parts. The three electric functions are 
\begin{align}
L_1 &= h_1  = 1 + \frac{Q_1}{8(1+b^2)}\left(\frac{1-b}{1+b}\frac{1}{|\vec x - \vec x_+|} + \frac{1+b}{1-b}\frac{1}{|\vec x - \vec x_-|} \right),  \\
L_2 &= h_2 =   1 + \frac{Q_2}{8(1+b^2)}\left(\frac{1-b}{1+b}\frac{1}{|\vec x - \vec x_+|} + \frac{1+b}{1-b}\frac{1}{|\vec x - \vec x_-|} \right), \\
L_3 &= 1,
\end{align}
and finally the function $M$ is,
\be
M = m_0 + \frac{m_+}{| \vec x - \vec x_+|} + \frac{m_-}{| \vec x - \vec x_-|},
\ee
where
\bea
m_0 &=& - \frac{\sqrt{Q_1 Q_2}}{4 \sqrt{2}(1-b^2)R} ,\\
m_+ &=& \frac{ R \sqrt{Q_1 Q_2}(1-b)^2}{32
   \sqrt{2} (1+b) \left(1+ b^2\right)^2} \left(1-b- i (1+b)  \right),\\
m_- &=& \frac{ R \sqrt{Q_1 Q_2}(1+b)^2}{32
   \sqrt{2} (1-b) \left(1+ b^2\right)^2} \left(1+b + i (1-b)  \right).
\eea
The coefficients of the function $M$ also have imaginary parts. 

Several comments are in order here. 
\begin{enumerate}
\item The magnetic function $K^3$ is dipolar: it is dipolar in the sense that it captures the dipole charge and also dipolar in the sense that the charges at the two centers are complex conjugate of each other. 
\item The electric functions are real. However, the charges are split between the $\vec x_\pm$ centers in a non-trivial (un-equal) way. The sum of charges at the two centers equals the total charge. 
\item The coefficients at the centers for the function  $M$ are fairly cumbersome. This is essentially due to the cumbersome nature of the Pomeransky-Sen'kov solution. The coefficients at the centers are not complex conjugate of each other.  
\end{enumerate}
It will be good to understand if these coefficients can be understood from the new form of attraction point of view \cite{Boruch:2023gfn}. Since the full solution has three centers, the formalism of \cite{Boruch:2023gfn} needs to be adapted somewhat to address this question. A further analysis of the solution, including the nature of the singularities at the centers, and its relation to other solutions discussed in the literature\footnote{Perhaps to the bubbling supertubes \cite{Bena:2005va} via the so-called spectral flow transformations \cite{Bena:2008wt} together with some analytic continuation.} is left for the future.

In the $b\to 0$ limit, the eight harmonic functions simply reduce to the eight harmonic functions for the two-charge black ring, $V = 1/|\vec x|, K^1= 0, K^2 = 0, L_3 =1$,
\begin{align}
M &=- \frac{\sqrt{Q_1 Q_2}}{4 \sqrt{2}R} + \frac{ R \sqrt{Q_1 Q_2}}{16
   \sqrt{2}|\vec x - \vec x_0|} , & K^3 & =\frac{\sqrt{Q_1 Q_2}}{2 \sqrt{2}R |\vec x - \vec x_0|},\\
L_1 &= 1 + \frac{Q_1}{4 |\vec x - \vec x_0|}, & L_2 &= 1 + \frac{Q_2}{4 |\vec x - \vec x_0|},
\end{align}
where $\vec x_0 = \left(0,0,-\frac{1}{4}R^2\right)$. 

\section{Conclusions and outlook}
\label{sec:conclusions}
Index saddles for small black holes are now reasonably well understood \cite{Chowdhury:2024ngg, Chen:2024gmc, Hegde:2024bmb}. It is natural to ask if those calculations can be extended to the case of small black rings? In this paper, we have taken a few steps in this direction. 

We  considered the supersymmetric two charge black ring of Elvang and Emparan \cite{Elvang:2003mj} and found a saddle  that contributes to the index for this black ring. Our main reason for working with the Elvang-Emparan solution is the  technical simplicity of the corresponding saddle solution, which rotates in both planes. It is certainly more interesting  to work 
with a supersymmetric black ring with one independent dipole charge. However, working with dipole black rings with both rotations is fairly cumbersome. This is because adding dipole charges \cite{Emparan:2004wy} on neutral black rings is not straightforward; for example, these charges cannot be added using boost-duality based solution generating techniques.

A versatile method is, however, known that adds a single dipole charge \cite{Rocha:2011vv} on neutral black rings. This method was used in, for example, ref.~\cite{Chen:2012kd},  to present solutions with both rotations and a dipole charge. Given the cumbersome nature of the metrics not much has been done using these solutions. However, we believe that these solutions with the addition of two charges can be used to construct  index saddle solutions for five-dimensional small black rings of \cite{Elvang:2004xi, Iizuka:2005uv, Dabholkar:2005qs, Dabholkar:2006za}. We hope to report on this in our future work \cite{to_apear}.

In the last few months, there has been significant progress in  understanding the nature of the index saddles for a variety of black holes in  five dimensions \cite{Boruch:2025qdq,Adhikari:2024zif,Cassani:2024kjn}. Certain ideas were discussed in \cite{Boruch:2025qdq} for constructing saddles for supersymmetric black rings. 
They suggested considering uplift of 4D multi-center supersymmetric configurations  to 5D and searching for saddles corresponding to black rings in this parameter space.  At present it is not clear to us how practical this method is. On the one hand, for the small black ring of Elvang and Emparan, which in our opinion is the simplest supersymmetric black ring set-up,  we could not have guessed the three-center solution in the form we reported in this paper. On the other hand, the non-extremal solutions with two rotations, three independent dipole,  and three independent electric charges are prohibitively cumbersome to work with. Perhaps a midway path is most likely to achieve the most success. Taking inspiration from the analysis we have performed, it is perhaps possible to find  saddles for finite entropy supersymmetric black rings by considering uplift of 4D multi-center supersymmetric configurations.  We hope to make progress on this question in our future work.

 \bigskip
 
\noindent \textbf{Acknowledgement:}  A.V. thanks Roberto Emparan and Ashoke Sen for  discussions that helped formulate the project. We would like to thank P Shanmugapriya and Pavan Dharanipragada for discussions and collaboration during the initial stages of this work. G.S.P. would like to thank Imtak Jeon and Robert de Mello Koch for organizing the visit, the local support, and the warm hospitality extended in Huzhou during the final stages of this work. A.V. thanks NISER Bhubaneswar for warm hospitality. The work of A.V. was partly supported by SERB Core Research Grant CRG/2023/000545.  We also thank Roberto Emparan and Ashoke Sen  for their comments on an earlier draft.

 \appendix

\section{Generating two charge solutions}
\label{app:solution_generating_technique}

We are interested in solutions to the classical equations of motion obtained from the action of the low energy NS-NS sector of the superstring theory compactified on $T^4 \times S^1$. It is convenient to start in six dimensions. A suitable truncation is to six-dimensional metric, antisymmetric two-form field $B_{MN}$, and dilaton $\Phi$. The six-dimensional action in string frame takes the form, 
 \be
 S_{6S} = \frac{1}{16\pi G_6} \int d^6x \sqrt{-G^{(S)}} e^{-2\Phi}\left[ R^{(S)} + 4(\nabla \Phi)^2  - \frac{1}{12}  H_{MNP}H^{MNP}\right], \label{action_6sd}
 \ee
 where $H = dB$.  We can go to Einstein frame using, 
\be
 G_{MN}^{(E)} = e^{-\Phi} G^{(S)}_{MN}.
\ee 
The Einstein frame action reads, 
\be
 S_{6E} = \frac{1}{16\pi G_6} \int d^6x \sqrt{-G^{(E)}} \left[ R^{(E)} -(\nabla \Phi)^2  - \frac{1}{12}  e^{-2\Phi}H_{MNP}H^{MNP}\right]. \label{action_6ed}
 \ee

A Kaluza-Klein reduction on a circle gives the theory of interest in five dimensions. Using the following ansatz for the metric,
 \be
ds^2_{6E} = e^{\frac{1}{\sqrt{6}}\chi} ds_{5E}^2 + e^{-\frac{\sqrt{3}}{\sqrt{2}} \chi} (dz + A^{(1)})^2,
\ee
we go to the five-dimensional Einstein frame. The ansatz for the reduction  of the NS-NS two-form $B(x,z)$ field is
\be
B(x,z) = B(x) + A^{(2)}(x) \wedge dz,
\ee 
where $B(x)$ is a two form in five dimensions and $A^{(2)}(x)$ is a one form.   We get the following five dimensional Einstein frame action upon dimensional reduction,
\be
 S = \frac{1}{16\pi G_N} \int d^5 x \sqrt{-g}~{\cal L}, \label{action_5d} \\
\ee
where
\be
{\cal L} = R - \frac{1}{2}(\nabla \chi)^2   - (\nabla \Phi)^2 - \frac{1}{12}e^{-\frac{\sqrt{2}}{\sqrt{3}}\chi - 2\Phi} H^2 -\frac{1}{4}e^{-\frac{2 \sqrt{2}}{\sqrt{3}}\chi}\left(F^{(1)}\right)^2
-\frac{1}{4}e^{\frac{\sqrt{2}}{\sqrt{3}}\chi -2\Phi}\left(F^{(2)}\right)^2,
 \ee
with the field strengths defined as
\be
H = d B - d A^{(2)} \wedge A^{(1)} \ ,
\ee
and $F^{(1)}= d A^{(1)}$,  $F^{(2)} = d A^{(2)}.$ 
 
 To construct charged solutions of interest to theory \eqref{action_5d}, we start with a general stationary metric that solves vacuum Einstein equations in five dimensions and generate a two charge metric using boosts and T-duality. We take the seed metric of the general form,
\be
ds^2_{5} = g_{rr} dr^2 + g_{\theta\theta}d\theta^2  + g_{\phi\phi}d\phi^2 + g_{\psi\psi}d\psi^2 + g_{tt}dt^2 + 2g_{\phi\psi}d\phi d\psi + 2g_{\phi t}d\phi dt + 2g_{\psi t}dt d\psi . \label{metric_5d_seed}
\ee
The five coordinates are $(r, \theta, \phi, \psi, t)$. Metric \eqref{metric_5d_seed} is diagonal in the $(r, \theta)$ space and has the most general form in the $(\phi, \psi, t)$ space, with $\partial_t, \partial_\phi, \partial_\psi$ as three commuting Killing vectors. For black rings, $r$ is to thought of as $y$ and $\theta$ as $x$.  All seed metrics that we use will be of this form. We can interpret metric \eqref{metric_5d_seed} as a string frame metric to ten-dimensional  string theory by adding five flat directions $z, z_1,z_2,z_3,z_4$ with all other 
  NS-NS and R-R fields set to zero. We have,
\bea
ds^2 &=& ds^2_{5} + dz^2 + \sum_{i=1}^{4}dz_i dz_i,  \\
e^{2\Phi} &=&1 ,  \qquad  B=0.
\eea
Since nothing happens in the $z_i$ directions, we ignore them in what follows and concentrate on the six-dimensional metric and fields. To add charges, we proceed in three steps:
\begin{enumerate}
\item We first perform a boost along the $z$-direction,
\bea
t&=&  t' \cosh \alpha +  z' \sinh\alpha,   \\ 
z&=&  z'  \cosh \alpha +  t' \sinh\alpha.
\eea
\item  Next we perform a T-duality along the $z'$ direction using the following rules (see e.g.,~\cite{johnson}). For ease of notation let us call $z'=s$. We have, the transformed fields as,
 \begin{align}
 G'_{ss}&=\frac{1}{G_{ss}}, &
 e^{2\Phi'}&=\frac{e^{2\Phi}}{G_{ss}}, \\ 
 G'_{\mu s}&=\frac{B_{\mu
 s}}{G_{ss}},& B'_{\mu s}&=\frac{G_{\mu s}}{G_{ss}},\\
G'_{\mu \nu}&=G_{\mu \nu}-\frac{G_{\mu s}G_{\nu s}-B_{\mu s}B_{\nu  
s}}{G_{ss}}, &
B'_{\mu \nu}&=B_{\mu \nu}-\frac{B_{\mu s}G_{\nu s}-G_{\mu s}B_{\nu  
s}}{G_{ss}}.
 \end{align}
\item Finally, we perform another boost in the   $z'$ direction with a different boost parameter,
\bea
t' &=& t'' \cosh \beta  + z''  \sinh\beta, \\ 
 z'&=&  z''  \cosh \beta  + t''  \sinh\beta.
\eea
\end{enumerate}

At the end of these steps, we get the desired fields in 6D in the string frame. Dimensionally reducing to 5D using the reduction procedure discussed above, we get the final 5D fields. Dropping the primes, we have the final Einstein frame metric $g^{E\: 5D}_{\mu \nu}$ in terms of the seed metric $g_{\mu \nu}$ as,
\bea
g^{E\: 5D}_{r r} &=& H g_{rr}, \\
g^{E\: 5D}_{\theta \theta} &=&  H  g_{\theta \theta},  
\eea
\bea
g^{E\: 5D}_{tt} &=& H^{-2} g_{tt}, \\
g^{E\: 5D}_{t \phi} &=& H^{-2} c_\alpha c_\beta g_{t\phi}, \\
g^{E\: 5D}_{t \psi} &=& H^{-2} c_\alpha c_\beta g_{t\psi}, 
\eea
\bea
g^{E\: 5D}_{\psi \psi} &=& H^{-2} \left( c_\alpha^2 c_\beta^2 g_{\psi \psi}+ (c_\alpha ^2 s_\beta^2 + s_\alpha^2 c_\beta^2+ s_\alpha^2 s_ \beta^2 g_{tt}) (g_{\psi\psi}g_{tt} -g^{2}_{\psi t})\right),\\
g^{E\: 5D}_{\phi \phi} &=& H^{-2}\left( c_\alpha^2 c_\beta^2 g_{\phi \phi}+ (c_\alpha ^2 s_\beta^2 + s_\alpha^2 c_\beta^2+ s_\alpha^2 s_ \beta^2 g_{tt}) (g_{\phi\phi}g_{tt} -g^{2}_{\phi t})\right), \\
g^{E\: 5D}_{\phi \psi} &=& H^{-2}\left(c_\alpha^2 c_\beta^2 g_{\psi \phi}+ (c_\alpha ^2 s_\beta^2 + s_\alpha^2 c_\beta^2+ s_\alpha^2 s_ \beta^2 g_{tt}) (g_{\psi\phi}g_{tt} -g_{\phi t}g_{\psi t})\right), 
\eea
where we use the short hand notation $c_\alpha = \cosh \alpha, c_\beta = \cosh \beta, s_\alpha = \sinh \alpha, s_\beta = \sinh \beta$ and  we have also defined,
\begin{align}
H & = h_\alpha^\frac{1}{3}h_\beta^\frac{1}{3}, &
h_\alpha &= c_\alpha^2 + s_\alpha^2 g_{tt}, &
h_\beta &= c_\beta^2 + s_\beta^2 g_{tt}.
\end{align}
The five-dimensional dilaton and the scalar $\chi$ take the form,
\begin{align}
e^{-2\Phi} & = h_\alpha, & e^{-\frac{\sqrt{3}}{\sqrt{2}} \chi}  &= \frac{h_\beta}{\sqrt{h_\alpha}}.
\end{align}
The two vectors $A^{(1)}$ and $A^{(2)}$ take the form,
\begin{align}
A^{(1)}_t &= \frac{(1 + g_{tt}) }{c_\beta^2 + g_{tt}s_\beta^2 } c_\beta s_\beta,  &
A^{(2)}_t &= \frac{(1 + g_{tt}) }{c_\alpha^2 + g_{tt}s_\alpha^2 } s_\alpha c_\alpha, \\  
A^{(1)}_\psi &= \frac{g_{t\psi} }{c^2_\beta + g_{tt}s^2_\beta } c_\alpha s_\beta, &
A^{(2)}_\psi &= \frac{g_{t\psi} }{c^2_\alpha + g_{tt}s^2_\alpha } s_\alpha c_\beta, \\ 
A^{(1)}_\phi &= \frac{g_{t\phi} }{c^2_\beta + g_{tt}s^2_\beta } c_\alpha s_\beta, & 
A^{(2)}_\phi &= \frac{g_{t\phi} }{c^2_\alpha + g_{tt}s^2_\alpha } s_\alpha c_\beta. 
\end{align}
Finally, the five-dimensional $B$-field has non-zero components as,
\bea
B_{\phi t} &=& \frac{g_{t\phi}s_\alpha s_\beta }{c^2_\alpha + g_{tt}s^2_\alpha } , \\ B_{\psi t} &=& \frac{g_{t\psi}s_\alpha s_\beta }{c^2_\alpha +g_{tt} s^2_\alpha }.
\eea

\paragraph{A five-dimensional duality relation: } We can convert the two-form potential $B$ into a one form potential $A^{(3)}$ and interpret the solution as a solution of U(1)$^3$ supergravity. The duality relation is, see for example \cite{Virmani:2012kw}\footnote{In \cite{Virmani:2012kw} the sign of the 5D Chern-Simons term was taken to be $+$, whereas in the Bena-Warner literature it is taken to be $-$, hence the sign difference.},
\be
\exp\left[ - \sqrt{\frac{2}{3}} \chi - 2 \Phi \right]\star_5 H = - d A^{(3)}.
\ee
Our conventions for the orientation with the black ring coordinates is $\epsilon_{t y x \phi \psi} = + \sqrt{-\det g}$. For the Hodge star operation we always use the Polchinski conventions \cite{Polchinski:1998rr}. 

\paragraph{Area computation:} To compute the area of a black hole we need the determinant of the five-dimensional Einstein frame metric on the section of constant $t$ and constant $r$. For singly spinning cases, with spin in the $\psi$ direction, $g_{t\phi} = 0$ and $g_{\phi \psi} = 0$. The determinant expression simplifies and we get,
\bea
(\det g^{E\: 5D})_{\theta \phi \psi} &=& c_\alpha^2 c_\beta^2 g_{ \theta\theta} g_{ \phi\phi}g_{ \psi\psi} \nonumber \\
& & + (c_\alpha^2 s_\beta^2 + s_\alpha^2 c_\beta^2) g_{ \theta\theta} g_{ \phi\phi} \left(g_{ \psi\psi} g_{tt} - g_{t\psi}^2 \right) \nonumber \\
& & + s_\alpha^2 s_\beta^2 g_{ \theta\theta}  g_{ \phi\phi} g_{tt}\left(g_{ \psi\psi} g_{tt} - g_{t\psi}^2 \right).
\eea
For the general situation define,
\be
g_3 = (\det g)_{\phi \psi t}  = 
\begin{pmatrix}
 g_{\phi \phi}  & g_{\phi \psi} & g_{\phi t}\\
g_{\phi \psi}  & g_{\psi \psi}  & g_{\psi t} \\
 g_{\phi t} & g_{\psi t}   & g_{tt}   
\end{pmatrix}.	   
\ee
We have,
\bea
(\det g^{E\: 5D})_{\theta \phi \psi} &=& c_\alpha^2 c_\beta^2 g_{\theta \theta} (g_{\phi \phi} g_{\psi \psi} - g_{\phi \psi}^2)  + (c_\alpha^2 s_\beta^2 + s_\alpha^2 c_\beta^2) g_{\theta \theta} g_3  + s_\alpha^2 s_\beta^2 g_{\theta \theta} g_{tt} g_3 .
\eea

\section{Saddle solution in the chiral null model form}
\label{app:chiral-null-model}
Taking motivation from \cite{Lunin:2001fv}, in this appendix we write the saddle solution uplifted to six-dimensions in the chiral null model form. Since all supersymmetric F1-P solutions can be written in this form, it is perhaps not a surprise that the saddle solution can be written in this form. Nonetheless, exhibiting this is interesting. It gives us confidence that saddle solutions for the higher dimensional  supersymmetric small black rings \cite{Dabholkar:2006za} can  be explored following this line of thought.

Recall that the five-dimensional vector fields and the B-field for the saddle solution take the simplified form given in \eqref{vector-saddle-1}--\eqref{vector-saddle-2} and \eqref{B-saddle}. Using these expressions,  we uplift the solution to six-dimensions and express it in string frame.  The 
string  frame metric reads, 
\bea
ds^2_{6S} &=& e^{\Phi} ds^2_{6E} = \frac{1}{\sqrt{h_1}}ds^2_{6E} \\
&=& -\frac{1}{h_1 h_2}\left( dt  + \Omega^{(s)} \right)^2   + ds^{2}_{\mathrm{flat}} + \frac{h_2}{h_1}(dz + A^{(1)})^2.
\eea
From the 5D $B$-field, we get the the following 6D $B$-field,
\be
B = B_{\psi t} d\psi \wedge dt  + B_{\phi t} d\phi \wedge dt  + A^{(2)} \wedge dz.
\ee

We can connect this form of the string frame metric to the chiral null model metrics \cite{Horowitz:1994rf,Tseytlin:1996yb}.  A standard form of the chiral null model is \cite{Lunin:2001fv},
\be
ds^2 = H\left(-dudv + K dv^2 + 2 {\cal A}_i dx^i dv \right) + dx_i dx_i,
\ee
with other fields given as,
\begin{align}
e^{-2\Phi} &= H^{-1},  &  B_{uv} &=  - \frac{1}{2}H, &   B_{vi} &= - H{\cal A}_i.
\end{align}
We can regard ${\cal A}_i$ as a vector field on the four-dimensional flat base space $dx_i dx_i$ and we construct ${\cal F}_{ij} = \partial_i {\cal A}_j - \partial_j {\cal A}_i$. Then, the functions in the chiral null model must  satisfy,
\begin{align}
\partial^2 H^{-1} &= 0, & \partial^2 K &= 0, & \partial_i {\cal F}^{ij} &=0.
\end{align}
In these equations, $\partial^2$ is the Laplacian in the $x_i$ Cartesian coordinates. 

From the Bena-Warner analysis of section \ref{sec:BW}, we conclude that the one form $\Omega^{(s)}$ on the four-dimensional flat base space satisfies,
\be
d\left[d \Omega^{(s)} + \star_4 d \Omega^{(s)}\right] = 0 \implies  d \star_4 d \left( \Omega^{(s)} \right) =0.
\ee
In particular, if we identify, ${\cal A}= - \Omega^{(s)},$ the chiral null model equations for ${\cal A}$ are satisfied. With the identifications 
 $H = h_1^{-1} , K = h_2 - 1, u = t-z$, and $v= t+ z$, the string frame metric for the saddle solution readily matches the chiral null model form. The other fields also match.

\end{document}